\documentstyle[aps,multicol,psfig]{revtex}
\begin{document}
\draft 

\title{Rolling friction of a hard cylinder on a viscous plane}

\author{Thorsten P\"oschel$^{1}$, Thomas Schwager$^{1}$, and Nikolai
  V. Brilliantov$^{1,2}$}

\address{$^1$Humboldt-Universit\"at zu Berlin, Institut f\"ur Physik,
  Invalidenstra{\ss}e 110, \\ D-10115 Berlin, Germany}

\address{$^2$Moscow State University, Physics Department, Moscow
  119899, Russia}

\date{\today}
\maketitle
\begin{abstract}
  The resistance against rolling of a rigid cylinder on a flat viscous
  surface is investigated. We found that the rolling-friction
  coefficient reveals strongly non-linear dependence on the cylinder's
  velocity. For low velocity the rolling-friction coefficient rises
  with velocity due to increasing deformation rate of the surface. For
  larger velocity, however, it decreases with velocity according to
  decreasing contact area and deformation of the surface.
\end{abstract}
\begin{multicols}{2}

\section{Introduction}
\label{sec:intro}
The effect of rolling friction has been investigated by many
scientists according to its great importance in engineering
(e.g.~\cite{Czichos:1978}) and physical science
(e.g.~\cite{Reynolds:1874}). Scientific publications on rolling
friction range back to, at least, 1785 when Vince described systematic
experiments to determine the nature of friction
laws~\cite{Vince:1785}.

It is known that surface effects such as adhesion
(e.g.~\cite{BarquinsEtAl:1978}), electrostatic interaction
(e.g.~\cite{DeryaguinSmilga:1994}), and other surface properties
(e.g.~\cite{ChaplinChilson:1986}) may have strong influence on rolling
friction. For viscoelastic materials, however, it was argued that
rolling friction is due very little to surface interactions, i.e. the
major part is due to deformation losses within the bulk of the
material~\cite{GreenwoodMinshallTabor:1961,Tabor:1955}. Under this
assumption Greenwood et al.~\cite{GreenwoodMinshallTabor:1961}
calculated the rolling friction coefficient for a hard sphere rolling
on a soft plane. The deformation in the bulk was assumed to be
completely plastic. Then an empirical coefficient was introduced to
account for the incomplete recover of the material. Recently a similar
problem has been addressed in~\cite{BrilliantovPoeschel:1998a} where
the rolling friction coefficient for a soft sphere on a hard plane has
been derived as a first-principle continuum-mechanics expression. This
coefficient has been found within a quasi-static
approach~\cite{BrilliantovSpahnHertzschPoeschel:1994} as a function of
the viscous and elastic constants of the sphere material without
introducing phenomenological parameters.

In the case of a soft sphere rolling on a hard
plane~\cite{BrilliantovPoeschel:1998a} the contact surface between the
bodies is flat. This allows for the application of Hertz's contact
theory. In the opposite case of a hard sphere or cylinder on a viscous
plane which we address here, this assumption is not justified since
the plane deforms in such a way that its shape follows the shape of
the rolling body in the area of contact.  This complicated shape of
the contact surface excludes the direct application of Hertz's contact
theory and may violate Hertz's contact law which relates the force
acting between the interacting bodies to their deformation.

The velocity dependence of the rolling friction coefficient originates
from the fact that the deformation of the surface varies with the
velocity of the rolling body. For small velocities the viscous stress,
proportional to the deformation rate, is small. In this case the
deformation of the plane (measured by a depth $h$ of which the body
penetrates the surface, see Fig.~\ref{fig:sketch}) is determined
mainly by the elasticity of the plane and by the weight of the body.
On the other hand at very large velocities the viscous stress becomes
comparable to the elastic stress. As a result the plane supports the
rolling body at significantly smaller deformations. This leads to
decreasing penetration depth $h$, hence, less energy may be required
to deform the surface. In this case one observes decreasing resistance
to rolling with increasing velocity.  Mainly because of the
complicatedly shaped contact area it is not possible to treat these
effects within the first-principle continuum-mechanics description,
hence a simpler model will be considered. As shown in the following
this model reflects the most important properties of the problem
addressed.
  \begin{minipage}{8cm}
\begin{figure}[htbp]
    \centerline{\psfig{figure=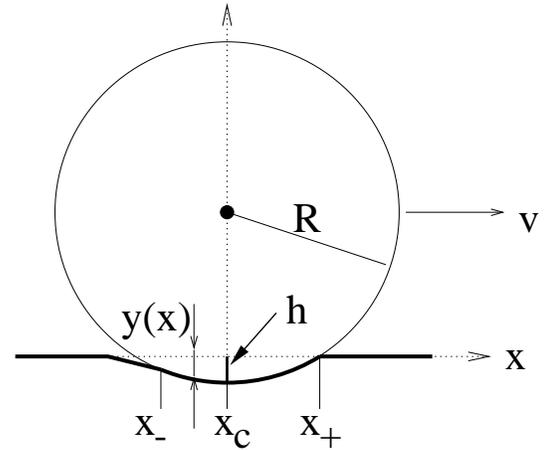,width=7cm,angle=0}}
\vspace{0.5cm}
  \caption{Sketch of the system: A rigid cylinder rolls on a plane 
    built up of independently moving damped springs. According to the
    motion of the cylinder the surface is deformed and mechanical
    energy is lost due to its damped motion. This dissipation of
    energy can be understood as rolling friction.}
  \label{fig:sketch}
\end{figure}
  \end{minipage}

\section{The Model}
\label{sec:model}
We investigate the resistance against rolling of a hard cylinder on a
soft plane.  Consider a cylinder of radius $R$, length $L \gg R$ and
mass per unit length $M$ which rolls along the $x$-axis with velocity
$v$.  We assume that the surface may be modelled by a system of
noninteracting springs. Their elastic, damping and inertial properties
are described by the coefficients $k$, $\gamma$ and $m$.  These are
defined as follows: $k\, dx$ and $\gamma \, dx$ give the elastic and
viscous force per unit length along the cylinder axis, $m \, dx$ gives
correspondingly the linear mass density of the springs (i.e. their
mass per unit length along this axis).  The viscous plane becomes
deformed in the range $x_- \leq x \leq x_+$ due to the mass of the
cylinder. For deformation rates small as compared with the vertical
speed of sound, Hooke's law is valid and one has the equation of
motion which describes the behaviour of the surface
\begin{equation}
  \label{eq:EqnOfMotion:Spring}
  m\ddot{y}(x)+\gamma\dot{y}(x)+ky(x)=f(x,t)\,,
\end{equation}
where $f$ is the force density ($f\, dx$ gives the force per unit
length along the cylinder axis) which acts on the plane in the region
of contact.  Outside of the contact area the force density is zero. In
our model we neglect lateral couplings of the springs (for
justification of these assumptions see the Appendix). We also assume
that no surface waves are excited on the plane, i.e. that the
condition of the overdamped motion of the surface
\begin{equation}
\label{dampcond}
\frac{m}{k} < \frac14 \frac{\gamma^2}{k^2}
\end{equation}
holds true~\cite{overdamped}.  Since we are interested in the steady
state, i.e. time independent behavior, we do not need to explicitly
consider the tangential interactions between the cylinder and the
surface. An arbitrarily small tangential force assures that the
cylinder does not slide. If one, however, is interested in accelerated
motion of the cylinder, tangential forces need to be considered.

By geometrical considerations we find for the shape $y(x)$ of the
surface of the deformed plane in the contact area:
\begin{equation}
  y(x) = R - h - \sqrt{R^2 -(x-x_{c})^2}\,,~~~ x_- \leq x \leq x_+\,,
\label{eq:y}
\end{equation}
where $x_c$ is the $x$-coordinate of the center and
$h=-y_{min}=-y(x_{c})$ is the penetration depth of the cylinder. For
$\left|x-x_c\right|\ll R$ we approximate (\ref{eq:y}) by
\begin{equation}
  \label{parabola}
  y(x)=\frac{(x-x_{c})^2}{2R}-h\,.
\end{equation}
The center of the cylinder moves with constant velocity $v$, i.e. $
x_{c}=vt$. Hence, the time derivatives of $y(x)$ read
\begin{eqnarray}
  \dot{y}(x)&=&-\dot{x}_{c}\frac{x-x_{c}}{R} = -v\frac{x-x_{c}}{R} 
  \label{parabola:td1}\\
  \ddot{y}(x)&=&\frac{v}{R}\dot{x}_{c} =\frac{v^2}{R}=\mbox{const}.  
  \label{parabola:td2}
\end{eqnarray}

The  compression force exerted by the plane to the cylinder is
\begin{equation}
  F_N = -\int \limits_{x_-}^{x_+} f(x) dx\,.
\end{equation}
(For simplicity of notation here and in what follows we notate the
forces, energy and torque per unit length of the cylinder, e.g. the
total force exerted by the plane to the cylinder is $L\,F_N$.)

The springs at $x_+$ which at time $t$ get in contact with the
cylinder need a separate discussion~\cite{singular}: At time
$t-\delta$ ($\delta \to 0$) their velocity is zero while infinitesimal
time later $\dot{y}(t+\delta)$ is finite according to
(\ref{parabola:td1}). This singularity in the velocity distribution
may be attributed to a force $F^\prime_N$, acting at point $x_+$. This
gives a finite contribution to the total force which can be determined
by the following consideration:

In the time interval $dt$ the cylinder moves by $vdt$. So it
accelerates springs of total mass $mvdt$. The total momentum received
by these springs is $dp=\dot{y}(x_+)\,mvdt$, hence,
\begin{equation}
  F^\prime_N = -\frac{dp}{dt} =-\dot{y}(x_+)m v =\frac{x_+-x_{c}}{R}\,mv^2\,.
\end{equation}
The total force $F_N + F^\prime_N$ supports thus the weight of the cylinder:
\begin{equation}
  F_N+F^\prime_N=Mg\,.
  \label{equilibrium}
\end{equation}

Substituting $y$ and its time derivatives
(Eqs.~(\ref{parabola},\ref{parabola:td1},\ref{parabola:td2})) in
(\ref{eq:EqnOfMotion:Spring}) we get an expression for the force
density in the contact area
\begin{equation}
  f(x,t)=\frac{k}{2R}\left(x-x_{c}\right)^2-\frac{\gamma v}{R}
\left(x-x_{c}\right)+\frac{m v^2}{R}-hk
  \label{force:x}
\end{equation}
which has to satisfy the contact condition
\begin{equation}
  f(x,t)\le 0~~~~\mbox{for}~~~x_- \leq x \leq x_+ 
\end{equation}
and which determines the boundaries of the contact area $x_\pm$. In
comoving coordinates $\xi=x-x_{c}$ Eq.~(\ref{force:x}) reads
\begin{equation}
  \label{force:xi}
  f(\xi)=\frac{k}{2R}\xi^2-\frac{\gamma v}{R}\xi+\frac{m v^2}{R}-hk\,.
\end{equation}
The boundary of the contact area at the front side of the cylinder in
the direction of motion is
\begin{equation}
  \label{eq:boundaryFront}
\xi_+=\sqrt{2Rh}  
\end{equation}
according to geometry. The boundary at the back side is determined by
$f(\xi_-)=0$, i.e.
\begin{eqnarray}
  0&=&\frac{k}{2R}\xi_-^2-\frac{\gamma v}{R}\xi_-+\frac{m v^2}{R}-hk\\
  \xi_-&=&\frac{\gamma v}{k} - \sqrt{2hR+\left(\frac{\gamma^2}{k^2}-2\,
      \frac{m}{k}\right) v^2}\,.
  \label{eq:boundaryBack}
\end{eqnarray}
For the self-consistency one needs the condition $\xi_-<\sqrt{2Rh}$,
i.e. $h(v) > m^2v^2/2\gamma^2R$ to be fulfilled, which restricts the
velocity $v$ from above~\cite{skiing}.

Because of the dissipative properties of the plane the motion of the
cylinder corresponds to a loss of mechanical energy per time. Another
contribution originates from the instantaneous acceleration of plane
material which gets in contact with the plane during the time
$dt$~\cite{singular}. Therefore, we find for the energy which is
transferred per time from the cylinder to the plane
\begin{eqnarray}
  \dot{E}&=&-\int\limits_{\xi_-}^{\xi_+}d\xi f(\xi)\dot{y}(\xi) 
  -m\frac{\dot{y}^2(\xi_+)}{2}v \nonumber \\
  &=&\frac{v}{R}\int\limits_{\xi_-}^{\xi_+}d\xi \xi f(\xi) 
  -mv^3\frac{2h}{R}
   \nonumber \\
  &=&-vF_R \,.
\label{FRdef}
\end{eqnarray}
Equation~(\ref{FRdef}) defines the force $F_R$ which acts against
rolling of the cylinder
\begin{equation}
  F_R=-\frac{1}{R}\int\limits_{\xi_-}^{\xi_+}d\xi \xi f(\xi)  + 
  m v^2\frac{2h}{R}\,.
\label{RF}
\end{equation}
To evaluate $F_R$ we need the force $f(\xi)$ given by
Eq.(\ref{force:xi}) with the penetration depth $h(v)$ which is to be
found. The penetration $h(v)$ results from equilibrating the dynamical
resistance of the surface with the weight of the cylinder
(\ref{equilibrium})
\begin{equation}
  Mg=-\int \limits_{\xi_-}^{\xi_+}d\xi\left[\frac{k}{2R}\xi^2-
    \frac{\gamma v}{R}\xi+\frac{mv^2}{R}-hk\right]+\frac{m v^2 }{R}
  \sqrt{2hR}
\label{hcalc}
\end{equation}
with the boundaries (\ref{eq:boundaryFront}) and
(\ref{eq:boundaryBack}). Equation~(\ref{hcalc}) is an implicit
equation for $h$.

We consider first the limit of small velocities.  In this limit
Eq.~(\ref{hcalc}) may be solved as a perturbation expansion,
$h=h^{(0)}+h^{(1)}v+h^{(2)}v^2 +\cdots$.  It is more convenient,
however, to solve (\ref{hcalc}) with respect to an expansion of
$\xi_+$:

\begin{equation}
\label{xi+}
\xi_+=\xi_{+}^{(0)}+v\xi_{+}^{(1)}+v^2\xi_{+}^{(2)}+\cdots
\end{equation}

Using Eqs.(\ref{eq:boundaryFront}), (\ref{eq:boundaryBack})
and (\ref{xi+}) one can further 
write the small-$v$ expansion for $\xi_-$:

\begin{equation}
\label{xi-}
\xi_-=-\xi_{+}^{(0)}+v\left(\frac{\gamma}{k}-\xi_{+}^{(1)} \right)
-v^2\left( \frac{\lambda }{2 \sqrt{2hR}}+\xi_{+}^{(2)} \right)+\cdots
\end{equation}
where 
$$
\lambda \equiv \left(\frac{\gamma^2}{k^2}-2\frac{m}{k} \right) \, . 
$$
Substituting Eqs.~(\ref{xi+},\ref{xi-}) into Eq.~(\ref{hcalc}) and
using (\ref{eq:boundaryFront}) one can solve it perturbatively to find
the front boundary

\begin{equation}
\label{1}
\xi_+=\left( \frac{3R}{2k} Mg \right)^{1/3}- 
\frac{ \lambda \, v^2}{4 \,\xi_{+}^{(0)}} +\cdots\,,
\end{equation}
where 
\begin{equation}
\xi_{+}^{(0)}=\left( \frac{3R}{2k} Mg \right)^{1/3}
\end{equation}
denotes the front boundary for the static case. Then from (\ref{xi-})
the rear boundary follows
\begin{equation}
\label{12}
\xi_-=-\xi_{+}^{(0)}+v\, \frac{\gamma}{k}- 
\frac{ \lambda \, v^2}{4 \, \xi_{+}^{(0)}} +\cdots\,.
\end{equation}
Correspondingly, the penetration depth reads:

\begin{equation}
\label{h1}
h(v) = h_0-\frac{\lambda \, v^2}{4 \, R} +\cdots
\end{equation} 
where 
\begin{equation}
h_0 \equiv h^{(0)} = \left(\xi_{+}^{(0)}\right)^2/2R  
\end{equation}
is the static penetration depth. From Eqs.~(\ref{h1}) and
(\ref{dampcond}) it follows that the penetration depth $h$ decreases
with increasing velocity.

Using the obtained expansions for $\xi_{+}$ and $\xi_{-}$ it is
straightforward to calculate the rolling friction force. Substituting
(\ref{force:x}), (\ref{1}) and (\ref{12}) into Eq.~(\ref{RF}) one
finally arrives at an
expression for  the rolling friction torque, ${\cal M}=R\,F_R$:

\begin{eqnarray}
\label{rol}
{\cal M}&=&\mu_{\rm roll} Mg \\
\mu_{\rm roll}&=&
\frac{\gamma}{k} \,  v  
-\frac{3 \lambda }{4 \xi_{+}^{(0)} } \, v^2 + \cdots 
\label{MU}
\end{eqnarray}

As it follows from Eq.~(\ref{MU}), in the limit of small velocities
the leading linear term depends only on the viscous and elastic
constants and does not depend on the inertial properties of the
material, characterized by $m$. This means that in this regime the
inertial effects in the deformation process of the plane are
negligible. The second nonlinear term takes into account via $m/k$ in
$\lambda$ the inertial effects up to ${\cal O}( v^2)$.  It is also
interesting to note, that while the linear term does not depend on the
radius of the body, the nonlinear term depends (via $\xi_{+}^{(0)}$)
on both, the radius of the cylinder and its mass per unit length $M$.

In the general case Eq.~(\ref{hcalc}) has to be solved numerically.
The velocity dependence of the penetration depth $h$ and of the
rolling friction coefficient $\mu_{\rm roll}$ are shown in
Fig.~\ref{HonV} and Fig.~\ref{MUonV}. Calculations were performed for
a steel cylinder of radius $R=0.1\, {\rm m}$, and of mass per unit
length $M=250 \, {\rm kg\, m^{-1}}$ rolling on the rubber surface with
the following material parameters: $m=100 \, {\rm kg \cdot m^{-2}}$,
$k=10^7 \, {\rm kg \cdot m^{-2} \cdot s^{-2}}$, $\gamma=5\cdot 10^5 \,
{\rm kg \cdot m^{-2} \cdot s^{-1}}$. These values were obtained from
density and elastic constants of rubber~\cite{Material}; the viscous
constant was estimated from the restitution coefficient for colliding
rubber spheres, similarly as in~\cite{BrilliantovPoeschel:1998a},
where the details of this estimate are given.

\begin{minipage}{8cm}
  \begin{figure}[htbp]
    \centerline{\psfig{figure=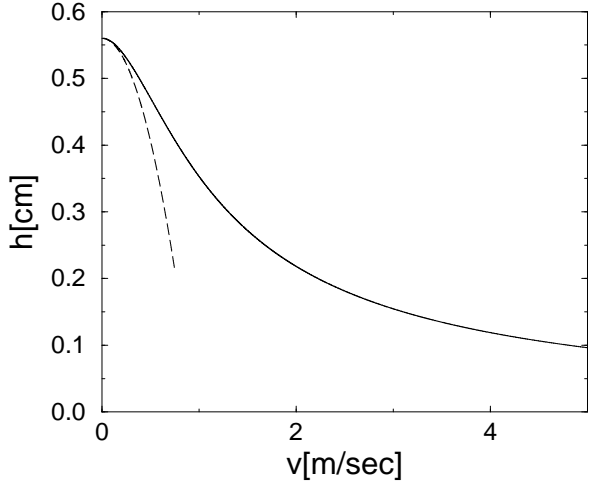,width=7.7cm,angle=0}}
    \caption{The penetration depth $h$ as a function of the velocity 
      $v$ according to the numerical solution of Eq.~(\ref{hcalc}).
      The dashed line shows the approximation~(\ref{h1}). }
    \label{HonV}
  \end{figure}
\end{minipage}

\begin{minipage}{8cm}
  \begin{figure}[htbp]
    \centerline{\psfig{figure=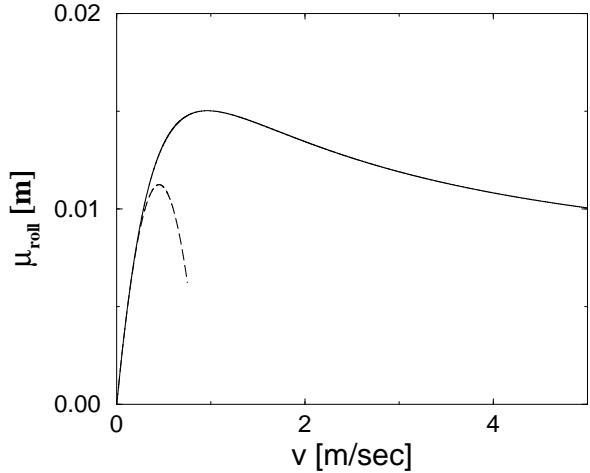,width=7.7cm,angle=0}}
    \caption{The rolling friction coefficient $\mu_{\rm roll}$ 
      over the velocity $v$ (numerical solution). The dashed line
      shows the approximation~(\ref{MU}).} \vspace{0.3cm}
    \label{MUonV}
  \end{figure}
\end{minipage}

As shown in the figures, at high velocities the depth $h$ decreases
and the rolling friction coefficient reaches a maximal value of
\begin{equation}
  \mu^{*}_{\rm roll}=\frac13 \, \frac {\gamma^2}{k^2}\, 
  \frac{\xi_{+}^{(0)}}{\lambda}
\end{equation}
at
\begin{equation}
  v^*= \frac23 \, \frac {\gamma}{k}\, \frac{\xi_{+}^{(0)}}{\lambda}\,.
\end{equation}
For velocities $v>v^*$ the rolling friction coefficient decreases with
increasing velocity. For particular parameters used here one obtains
$\mu^{*}_{\rm roll}=1.125 \cdot 10^{-2} \, {\rm m}$ and $v^*= 0.4500
\, {\rm m \, s^{-1}}$, which are in a reasonable agreement with the
numerically obtained values (see Figs.~\ref{HonV} and~\ref{MUonV}).

Note that with increasing velocity the rear boundary $\xi_-$ shifts in
positive direction and approaches the front boundary $\xi_+$.  The
contact area between the surface and the cylinder thus decreases and
at some critical velocity it shrinks to a point (a line along the
cylinder axis). For velocities larger than this critical one the
weight of the cylinder is sustained by the force $F^\prime_N$, acting
at a single point at the front boundary $\xi_+$.

\section{Summary and discussion}
\label{sec:Summary}

We investigated the rolling motion of a hard cylinder on a viscous
plane. The elastic, viscous and inertial properties of the plane were
modelled by a system of uncoupled springs which are characterized by
linear elastic and viscous coefficients and mass density.  For small
velocity of the rolling cylinder we determined the velocity expansion
of the rolling friction coefficient analytically up to second order.
For larger velocities the rolling friction was determined numerically.

In the range of low velocity our analysis shows increasing rolling
friction coefficient with increasing velocity. At a certain velocity
the coefficient reaches its maximal value and decreases when the
velocity is further increased.

In the low velocity regime where the rolling friction coefficient
rises linearly with the velocity of the cylinder its value depends on
the ratio of the viscous and elastic constants. For high velocities
one notes nonlinear dependence on the cylinder's mass, radius and on
the material constants of the surface.

We also analyzed the dependence of the penetration depth (i.e. the 
depth at which the cylinder sinks into the bulk of the surface) on the 
cylinder velocity. We found that the penetration depth decreases with 
increasing velocity, i.e. that the rolling cylinder emerges upwards 
when its velocity increases. 

\acknowledgments 
Frank Spahn is acknowledged for discussion. This work
was supported by Deutsche Forschungsgemeinschaft (Po 472/3-2 and
472/5-1).

\appendix
\section*{Lateral Couplings}
In our model we assumed that the viscous surface is composed of
springs which are not coupled in lateral (horizontal) direction. This
simplification may limit the validity of the model since we ignore
lateral interactions in the plane, which obviously exist in real
systems and which cause the tangential elasticity of the surface. In
the appendix we want to study in more detail the range of validity of
our model.

Lateral interactions of the springs may be taken into account if
instead of Eq.~(\ref{eq:EqnOfMotion:Spring}) one uses
\begin{equation}
  \label{eq:EqnOfMotion:Spring+d}
  m\ddot{y}(x)+\gamma\dot{y}(x)+ky(x)
  -d\frac{\partial^2}{\partial x^2}y=f(x,t)\,,
\end{equation}
where the constant $d$ describes the lateral coupling between the
springs. With the dimensionless variables $\hat{y}=y/R$, $\hat{x}=x/R$
and $\hat{t}= t/(R/v)$ the left-hand side of
Eq.~(\ref{eq:EqnOfMotion:Spring+d}) reads
\begin{equation}
  \label{eq:EqnOfMotion:Spring+ddimles}
  \frac{mv^2}{R}\frac{\partial^2}{\partial\,  \hat{t}^2} \hat{y}(\hat{x})
  +\gamma v \frac{\partial}{\partial \hat{t}}\hat{y}(\hat{x})
  +k R \hat{y}(\hat{x})-\frac{d}{R}\frac{\partial^2}{\partial \hat{x}^2}
  \hat{y}(\hat{x})\,.
\end{equation}
Thus we conclude that if the radius of the cylinder satisfies the
condition
\begin{equation}
  \label{cond}
  R^2 \gg \frac{d}{k} \, , 
\end{equation}
the term describing the lateral coupling is negligible and our model
is valid.

The value of $\sqrt{d/k}$ is a characteristic length of influence of
lateral couplings. If the above condition (\ref{cond}) is satisfied,
one can apply Eq.~(\ref{eq:EqnOfMotion:Spring}) for almost the entire
contact region, except for the (small) region around $\xi_{+}$. In
these region the deformation of the surface may differ from the
idealized shape which, as we assumed, results from pure geometrical
considerations. Instead the range of deformation of the surface is
slightly enhanced by a characteristic size $\sim r_0$ due to lateral
couplings between the springs. If the cylinder is at rest or if it
moves very slowly, similar discussion applies to the small region
$\sim r_0$ around the rear point $\xi_-$. Consider first the motion of
the surface in a region around $\xi_{+}$.

Since $\xi_{+} \gg r_0$ the region of the characteristic size $\sim
r_0$ may be considered as a point, so that the total force acting on
this region may be attributed to the single point $\xi_{+}$. The
dissipation in this region (which is finite) is, therefore, attributed
to the point $\xi_+$ too. Coarse-graining, therefore, results in the
force $F_N^{\prime}\left(\xi_{+}\right)$ acting at the point $\xi_+$.

We will estimate the characteristic size $r_0$: If we take into
account lateral couplings in the region $r_0$ around $\xi_+$ there is
no contact between the surface and the cylinder (free surface
condition). The derivative $y^{\prime} \equiv \partial y/ \partial x$
changes from zero (condition of the undisturbed plane) to $y^{\prime}
\approx y^{\prime} (\xi_{+}) \approx \xi_{+} /R$ at the point of
contact, which follows from the geometry of the system (see
Fig.~\ref{fig:sketch}). Thus, in this region one estimates
$y^{\prime\prime} \equiv \partial^2 y/ \partial x^2 \sim \xi_{+}
/R\,r_0$. Similarly, the characteristic value of $y$ in this region
reads $y \sim r_0 \cdot y^{\prime} \sim r_0 \cdot \xi_{+} /R$ and,
correspondingly, the characteristic values
$\dot{y}=\dot{y}(\xi_{+})=-v \xi_{+}/ R$ and $\ddot{y}=v^2/R$ follow
from Eqs.~(\ref{parabola:td1}), (\ref{parabola:td2}). Then we write
the condition of the free surface in this transient region,
\begin{equation}
  \label{eq:EqnOfMotion:Spring+dtransient}
  mv^2/R-\gamma v \xi_{+}/ R + k r_0 \xi_{+} /R -d \cdot 
  \xi_{+} /R\,r_0=0 \, , 
\end{equation} 
to estimate the size of the region:
\begin{equation}
  \label{r0}
  r_0=\sqrt{\frac{d}{k}+
    \left(\frac{mv^2}{2k \xi_{+} } -\frac{\gamma v}{2k} \right)^2 }-
  \left(\frac{mv^2}{2k \xi_{+} } -\frac{\gamma v}{2k} \right)
\end{equation} 
which yields $r_0 = \sqrt{d/k}$ for the static case. Hence, the
condition for coarse-graining, $r_0 \ll \xi_{+}$, reads
\begin{equation}
  \label{coarse}
  \frac{d}{k} \ll \xi_{+}^2 + \frac{m}{k}v^2 -\frac{\gamma}{k}v \xi_{+}\,.
\end{equation} 
We want to discuss the consequences of the assumption (\ref{coarse})
or of the assumption $\xi_{+}^2 \gg d/k $ (the later condition follows
from the former one, unless the velocity is too high, i.e.  unless $v
\gg \xi_{+} \sqrt{k/m}$). In this case the force $F_N^{\prime}
(\xi_{+})$ which acts at the point $\xi_{+}$ reads
\begin{equation}
  \label{impforce}
  F_N^{\prime} (\xi_{+}) =\int_{\xi_{+}-\delta}^{\xi_{+}+\delta} 
  f(\xi,t)d \xi \,,
\end{equation} 
where $\delta$ is of the order (say somewhat larger) than $r_0$, and
we can write for the different terms in the left-hand side of
Eq.~(\ref{eq:EqnOfMotion:Spring+d}):
\begin{eqnarray}
  \label{1term}
  &&\int_{\xi_{+}-\delta}^{\xi_{+}+\delta}m \ddot{y}dx= 
  m \int_{\xi_{+}-\delta}^{\xi_{+}+\delta} \frac{d \dot{y}}{dt} dx =
  m \int_{\xi_{+}-\delta}^{\xi_{+}+\delta} d \dot{y} \frac{dx}{dt}  
  \nonumber \\ 
  &&=mv \int_{\xi_{+}-\delta}^{\xi_{+}+\delta} d\dot{y} =
  mv \left[\dot{y}(\xi_{+}+\delta)- \dot{y}(\xi_{+}-\delta) \right] 
  \nonumber \\ 
  &&= -mv\dot{y}(\xi_{+})=mv^2 \,\xi_{+} /R 
\end{eqnarray}  
where we take into account that $\dot{y}(\xi_{+}+\delta)=0$ (the
surface is at rest) and that $\dot{y}(\xi_{+}-\delta) =
\dot{y}(\xi_{+})$ on the coarse-grained scale. Similarly, using the
above estimate of $y(x)$ in the transient region, we obtain the
coarse-grained estimates:
\begin{equation}
  \label{2term}
  \gamma \int_{\xi_{+}-\delta}^{\xi_{+}+\delta} \dot{y}dx= 
  \gamma v \left[y(\xi_{+}+\delta)- y(\xi_{+}-\delta) \right] \sim  
  -\gamma v r_0 \xi_{+} /R
\end{equation}
and  
\begin{equation}
  \label{3term}
  k \int_{\xi_{+}-\delta}^{\xi_{+}+\delta} y dx
  \approx k \int_{\xi_{+}-\delta}^{\xi_{+}+\delta} 
  y^{\prime}(\xi_{+})x dx
  \sim  2 k r_0  \xi_{+}^2 /R \, . 
\end{equation}
Finally, the last term reads
\begin{eqnarray}
  \label{4term}
  &&-d \cdot \int_{\xi_{+}-\delta}^{\xi_{+}+\delta} 
  y^{\prime \prime} dx 
  =-d \cdot \left[y^{\prime} (\xi_{+}+\delta)- 
    y^{\prime}(\xi_{+}-\delta) \right] \nonumber \\
  &&=d \cdot y^{\prime}(\xi_{+})=d \cdot \xi_{+}/R \,.
\end{eqnarray} 
As it follows from Eqs.~(\ref{1term}-\ref{4term}) the second and third
terms, proportional to $r_0$, vanish on the coarse-grained scale. The
fourth term does contribute to $F_N^{\prime} (\xi_{+})$ on the
coarse-grained level, but it does not depend on the velocity $v$. It
may be taken into account within the general scheme of calculation of
the rolling friction given above. Namely, with this term included, one
obtains, e.g. for the front boundary
\begin{equation}
\label{front+d}
\xi_{+}^{(0)}=\left\{
  \frac{3R\,Mg}{2k} \left[1+\frac32 
    \frac{ \left(d/k \right)}{\left( \xi_{+}^{(0)} \right)^2 } \right]
\right\}^{1/3}
\approx \left( \frac{3R\, Mg}{2k} \right)^{1/3}\, ,
\end{equation}
where the condition $\xi_{+}^2 \gg d/k $ was used. Similarly, the
impact of this term on the other expressions obtained previously is
negligible, i.e. it is of the order $(d/k)/
\left(\xi_{+}^{(0)}\right)^2 \ll 1$ under the coarse-grained
condition. Thus, we conclude that the fourth term in $F_N^{\prime}
(\xi_{+})$, which accounts for the lateral interactions may be also
neglected. This gives the result
\begin{equation}
  F_N^{\prime} (\xi_{+}) =mv^2 \, \xi_{+}/R\,.
\end{equation}

Similar considerations may be performed for the dissipation in the
transient region. Skipping the details of the analysis (very similar
to that for $F_N^{\prime} (\xi_{+})$), we give the final result:
\begin{equation}
  \label{dissimp}
  \int_{\xi_{+}-\delta}^{\xi_{+}+\delta} f(\xi,t)\dot{y}(\xi)d\xi 
  =-m \frac{\dot{y}^2(\xi_{+})}{2}v
\end{equation} 
which describes the energy loss in the point $\xi_{+}$ on the
coarse-grained scale.

Using the same reasoning one can consider the region $\sim r_0$ around
the rear point $\xi_-$ to conclude that under the condition $\xi_{+}^2
\gg d/k $ its contribution to the total force and to the dissipation
is negligible: Indeed, the region $\sim r_0$ in the rear part of the
contact area may influence the motion of the cylinder only for very
small velocities $v$ when $\left| \xi_+ \right|-\left| \xi_- \right|
\sim r_0$.  However, contribution to the force and dissipation from
this rear region is proportional to $r_0$ and therefore may be
neglected.

Thus, we conclude that our simplified model of the viscous surface as
a system of linear uncoupled springs may be adequate for real systems
and it may be used to model the rolling friction phenomenon.

\end{multicols}

\begin{references}

\bibitem{Czichos:1978}
H.~Czichos, {\em Tribology} Elsevier (Amsterdam, 1978);
R.~H. Bentall and K.~L. Johnson, Int. J. Mech. Sci. {\bf 6}, 389 (1967);
F.~T. Barwell, {\em Bearing Systems: Principles and Practice}, Oxford 
Univ. Press (Oxford, 1979), p.~312;
A.~Z. Szeri, {\em Tribology-Friction, Lubrication, and Wear}, 
Hemisphere (Washington, 1980), p.~401;
D.~Dowson, C.~M. Taylor, T.~H.~C. Childs, M.~Godet, and G.~Dalmaz 
(Eds.), {\em Dissipative Processes in Tribology}, Elsevier (Amsterdam, 1994);
H.~Krause and H.~Lehna, 
 {Wear} {\bf 119}, 153 (1987);
G.~P. Shpenkov, {\em Friction Surface Phenomena}, Elsevier (Amsterdam, 1995);
E.~Rabinowicz, {\em Friction and Waer of Materials}, Wiley (New York, 1965);
K.-H.~zum Gahr, {\em Microstructure and Wear of Material}, Elsevier 
(Amsterdam, 1987);
F.~W. Carter,
 {Proc. Roy. Soc. A} {\bf 112}, 151 (1926);
A.~Domen\'ech, T.~Domen\'ech, and J.~Cebri\'an,
 {Am. J. Phys.} {\bf 55}, 231 (1986);
K.~R. Eldredge and D.~Tabor, 
 {Proc. Roy. Soc. A} {\bf  229}, 180 (1955);
D.~G. Evseev, B.~M. Medvedev, G.~G. Grigoryan, and O.~A. Ermolin,
 {Wear} {\bf 167}, 33 (1993);
Y.~F. Li and A.~Seireg,
 {J. Tribilogy-Transactions of the ASME} {\bf 111}, 386 (1989);
S.~Lingard,
 {Wear} {\bf 117}, 109 (1987);
D.~E. Shaw and F.~J.~Wunderlich,
 {Am. J. Phys.} {\bf 52}, 997 (1984);
P.~H. Vo,
 {Proceedings of the SPIE - Int. Soc. Opt. Engin.} {\bf 1998}, 141 (1993);
K.~Weltner,
 {Am. J. Phys.} {\bf 55}, 937 (1987);
J.~Witters and D.~Duymelink,
 {Am. J. Phys.} {\bf 54}, 80 (1986).

\bibitem{Reynolds:1874}
O.~Reynolds,
 {Phil. Trans. Roy. Soc.} {\bf 166}, 1 (1874);
R.~Q. Huang and S.~T. Wang,
In: Chr. Bonnard (Ed.), {\em Proc. Intern Symp. on Landslides, 
Lausanne, Juli 10-15, 1988} (1988), p.~187;
H.~J.~Herrmann, G.~Mantica, and D.~Bessis,
 {Phys. Rev. Lett.} {\bf  65}, 3223 (1990).

\bibitem{Vince:1785}
S.~Vince.
 {Phil. Trans. Roy. Soc. London} {\bf 75}, 165 (1785).

\bibitem{BarquinsEtAl:1978}
M.~Barquins, D.~Maugis, J.~Blouet, and R.~Courtel,
 {Wear} {\bf 51}, 375 (1978);
B.~V. Deryaguin and Y.~P. Toporov,
 {Progr. in Surface Sci.} {\bf 45}, 317 (1994);
K.~Kendall,
 {\em Wear} {\bf 33}, 351 (1975);
K.~N.~G. Fuller and A.~D. Roberts,
 {J. Phys. D} {\bf 14}, 221 (1981);
A.~D. Roberts and A.~G. Thomas,
 {\em Wear} {\bf 33}, 45 (1975).

\bibitem{DeryaguinSmilga:1994}
B.~V. Deryaguin and V.~P. Smilga,
 {Progr. in Surface Sci.} {\bf 45}, 108 
and 296 (1994);
H.~Prashad,
 {J. Tribilogy-Transactions of the ASME} {\bf 110}, 448 (1988);
P.~V. Nazarenko, V.~F. Labunets, V.~S. Pilyavsky, and M.~N. Popovich,
 {Dopovidi Akademii Nauk Ukrainskoi RSR A} {\bf 2}, 68 (1986).

\bibitem{ChaplinChilson:1986}
R.~L. Chaplin and P.~B. Chilson,
 {Wear} {\bf 107}, 213 (1986);
Q-g. Song;
 {Am. J. Phys.} {\bf 56}, 1145 (1988);
R.~L. Chaplin and M.~G. Miller,
 {Am. J. Phys.} {\bf 52}, 1108 (1984).

\bibitem{GreenwoodMinshallTabor:1961}
J.~A. Greenwood, H.~Minshall, and D.~Tabor,
 {Proc. Roy. Soc. A} {\bf 259}, 480 (1961).

\bibitem{Tabor:1955}
D.~Tabor,
 {Proc. Roy. Soc. A} {\bf 229}, 198 (1955);
{\bf 43}, 1055 (1952);
D.~Atack and D.~Tabor,
{Proc. Roy. Soc. A} {\bf 246}, 539 (1958);
R.~C.~Drutowski,
In R.~Davies (Ed.), {\em Proc. Symp. Friction and Wear}, 
Elsevier (Amsterdam, 1959), p.~17.

\bibitem{BrilliantovPoeschel:1998a}
N.~V.~Brilliantov and T.~P\"oschel,
 {Europhys. Lett} {\bf 42}, 511 (1998).

\bibitem{BrilliantovSpahnHertzschPoeschel:1994}
N.~V. Brilliantov, F.~Spahn, J.-M.~Hertzsch, and T.~P\"oschel,
 {Phys. Rev. E} {\bf 53}, 5382 (1996).
 
\bibitem{overdamped} 
For the case of lower damping the situation
  appears to be more complicated: Surface waves of the plane are
  excited and it may happen that the area of contact is not
  continuous, i.e. the sphere and the surface touch in multiple
  regions $x^{(1)}_- \leq x \leq x_+^{(1)}$ and $x^{(2)}_- \leq x \leq
  x_+^{(2)} < x_-^{(1)}$, etc. This complicated behaviour which
  resembles Schallamach waves is excluded if we require overdamped
  motion of the plane.  Actually, it may be shown that more weak than
  Eq.~(\ref{dampcond}) condition, $m/k < \gamma^2/2k^2$, should be
  satisfied to exclude the multiple contact areas. This corresponds to
  surface waves appearing in the rear part of the plane which has been
  already passed by the rolling cylinder. Such kind of waves will not
  affect our calculations.
  
\bibitem{singular} 
In real systems where the surface also has a
  tangential elasticity, which may be modelled by the lateral coupling
  between the springs, one should consider a transient region of the
  plane of some finite extent. For the case of large radius of the
  cylinder, compared to the penetration depth, and for large vertical
  elasticity, compared to the tangential elasticity, the transient
  region is small. One can then attribute the force and dissipation
  occuring in this region to a single point.  Since the region is
  small, the viscous losses are negligible compared to the energy
  required to accelerate this part of the surface material. More
  details are given in the Appendix.
  
\bibitem{skiing} 
If this condition does not hold the cylinder is
  sustained only by the singular force $F^\prime_N$. This situation
  reminds to water skiing, when all the weight of the skier is
  sustained by the inertia of the water. In this case the lateral
  coupling of springs ignored in the present model is important.

\bibitem{Material}
Landolt-B\"ornstein.
 {\em New Series, vol.V/1b}.
 Springer Verlag, (Berlin, 1982).

\end{references}
\end{document}